\documentclass[fleqn,usenatbib]{mnras}
\usepackage{newtxtext,newtxmath}
\usepackage[T1]{fontenc}
\DeclareRobustCommand{\VAN}[3]{#2}
\let\VANthebibliography\thebibliography
\def\thebibliography{\DeclareRobustCommand{\VAN}[3]{##3}\VANthebibliography}
\usepackage{graphicx}
\usepackage[table,xcdraw]{xcolor}
\usepackage{graphicx}	
\usepackage{amsmath}	
\usepackage{pdflscape}
\usepackage{capt-of}
\usepackage{textcase}
\usepackage{afterpage}
\usepackage{longtable}
\usepackage{adjustbox}
\usepackage{placeins}
\title[Classical Ae stars in LAMOST DR5]{Identification of new Classical Ae stars in the Galaxy using LAMOST DR5}
 
\author[Anusha et al.]{R. Anusha,$^{1}$\thanks{E-mail: r.anusha@phy.christuniversity.in}
Blesson Mathew,$^{1}$
B. Shridharan,$^{1}$
R. Arun,$^{1}$
S. Nidhi,$^{1}$
Gourav Banerjee,$^{1}$
\newauthor{Sreeja S. Kartha,$^{1}$ K. T. Paul$^{1}$ and Suman Bhattacharyya$^{1}$}\\
$^{1}$Department of Physics and Electronics, CHRIST (Deemed to be University), Hosur Main Road, Bangalore, India\\}
\date{Accepted XXX. Received YYY; in original form ZZZ}

\pubyear{2020}

\begin{document}
\label{firstpage}
\maketitle

\begin{abstract}
We report the first systematic study to identify and characterize a sample of classical Ae stars in the Galaxy. The spectra of these stars were retrieved from the A-star catalog using the Large sky Area Multi-Object fiber Spectroscopic Telescope (LAMOST) survey. We identified the emission-line stars in this catalog from which 159 are confirmed as classical Ae stars. This increases the sample of known classical Ae stars by about nine times from the previously identified 21 stars. The evolutionary phase of classical Ae stars in this study is confirmed from the relatively small mid- and far-infrared excess and from their location in the optical color-magnitude diagram. We estimated the spectral type using MILES spectral templates and identified Classical Ae stars beyond A3, for the first time. The prominent emission lines in the spectra within the wavelength range 3700 -- 9000 {\AA} are identified and compared with the features present in classical Be stars. The H{$\alpha$} emission strength of the stars in our sample show a steady decrease from late-B type to Ae stars, suggesting that the disc size may be dependent on the spectral type. Interestingly, we noticed emission lines of Fe{\sc ii}, O{\sc i} and Paschen series in the spectrum of some classical Ae stars. These lines are supposed to fade out by late B-type and should not be present in Ae stars. Further studies, including spectra with better resolution, is needed to correlate these results with the rotation rates of classical Ae stars. 

\end{abstract}

\begin{keywords}
stars: emission-line, Be -- techniques: spectroscopic, photometric -- stars: circumstellar matter 
\end{keywords}
  
\section{Introduction}
A ‘Classical Be star' (CBe) was first defined by {\cite{1987Collins}} to designate non-supergiant B-type stars characterized by Balmer emission lines in their optical spectrum, at least once in their lifetime. This definition distinguishes them as a separate class from Be stars in the pre-main-sequence (PMS) phase, known as Herbig Be stars. A CBe star possesses a geometrically thin, gaseous, equatorial, decretion disc revolving in Keplerian rotation \citep{2007Meilland}. \cite{1931Struve} proposed that the formation of such a circumstellar envelope is due to the mass ejection episodes which occur as a result of the rapid rotation of CBe stars. But this disc formation mechanism in CBe stars, known as the ‘Be phenomenon’, still remains unclear. Subsequent works have provided clues that in addition to rapid rotation,  non-radial pulsations \cite[e.g.][]{1982Baade, 1998Rivinius, 2003Rivinius}, stellar wind \cite[e.g.][]{1993Bjorkman}, magnetic field \cite[e.g.][]{2002Cassinelli} and binarity \cite[e.g.][]{1975Kriz,2001Gies} are needed to understand the ‘Be phenomenon’ in CBe stars. A review of studies done till now in the field of CBe star research are provided by \cite{2013Rivinius} and \cite{2003Porter}. 

In addition to  CBe stars, the ‘Be phenomenon’ was identified in late O- and early A-type stars as well \citep{2004Negueruela}. Classical Ae (CAe) stars can be considered as late-type analogs to CBe stars, belonging to luminosity classes III--V and exhibiting a similar type of spectrum with emission lines as that of CBe stars \citep{1986Jaschek}. \cite{2005Cranmer} noted that late-type Be and early-type Ae stars are relatively at ease in forming a Keplerian disc because of their closer proximity to the critical rotation speed. This supports the view of successful angular momentum transfer in Ae stars even though they are relatively less active \citep{2017Baade}. Though studies have been performed on the disc characteristics and formation mechanisms of classical Oe stars \citep{1974Conti, 1976Frost, 2018Naze, Li2018}, a similar analysis has not been performed for CAe stars.

However, a few studies on infrared excess detected in CAe stars are worth mentioning \citep[e.g.][]{1998Jaschek, Jaschek_b1991, 1991Jaschek}. Flux excess in the infrared region (known as IR excess) observed in the spectral energy distribution of CBe stars originate due to the thermal bremsstrahlung emission from free electrons in the disc of CBe stars \citep[e.g.][]{1974Gehrz,2005Zhang,2015Vieira}. \cite{1991Jaschek} found that IR excess emission seen in CAe stars is similar to that of CBe stars. This suggests that the IR emission in CAe stars occur due to the presence of free electrons from the circumstellar disc. Further, \cite{2003Monin} did not find $V$-[12] color excess in Ae stars, $\nu$~Cyg, and $\kappa$~Ma, thus implying the absence of dust in the envelope of CAe stars. Hence, IR excess can be used to distinguish CAe stars from PMS stars such as Herbig Ae stars.

\cite{1986Andrillat} noticed that the frequency of Ae stars decrease as we move towards late spectral types. If the observed frequency distribution of CBe stars is continued towards early A-type stars, only 3\% of A1--A2 type dwarfs are expected to show Be-phenomenon \citep{2003Monin}. However, \cite{1997Zorec} identified only 0.2\% of Ae stars belonging to A1--A2 spectral types. In order to understand this number discrepancy, we checked {$\approx$ 40} CAe stars from the studies of \citet{1986Jaschek}, \citet{1997Kohou}, \citet{2003Monin}, \citet{2016Gkou} and \citet{2016Bohlender}. We found that many of these stars were later reported to be late B-type stars \citep{Sato1990, Anderson2012, 2015Raddi, 2019Liu}. This reduced the number of previously identified CAe stars to 21. The relatively small number of known CAe stars identified till now poses a formidable challenge in characterizing them as a separate group. This motivated us to search for a sample of CAe stars from the A-type emission-line stars identified from the LAMOST DR5 survey.

In this paper, we studied a sample of 2,931 A-type emission-line stars obtained from LAMOST DR5 and report the detection of 159 CAe stars from the analysis of spectra available from the same survey. Out of these 159 stars, only 58 are reported in literature to date. According to SIMBAD database, 55 objects are mentioned as ‘star', 2 are mentioned as white-dwarf candidates \citep{2013Zhang_WD} and another one is reported to be an young-stellar object \citep{2013Jose}. There exist no considerable information about the remaining 101 objects in the literature. 
Since  none of the 159 stars are reported as CAe in literature, they can be considered as new detections. This is the first systematic study performed till date to identify and characterize a sample of CAe stars from any database of emission-line stars. We also performed the first spectroscopic survey for a sample of CAe stars, thus providing an atlas of emission lines present in CAe stars for the first time. The paper is organized as follows. Sect. \ref{sec:Data} provides a brief introduction of the LAMOST survey. Sect. \ref{sec:Iden} explains the procedure and the criteria we followed for identifying the CAe stars. In Sect. \ref{sec:result}, we discuss the prominent emission features found in the spectra of our sample of CAe stars. The major results of our study are summarized in Sect. \ref{sec:concl}.  

\section{Data Inventory}
\label{sec:Data}

The Large sky Area Multi-Object fiber Spectroscopic Telescope (LAMOST) is a 4-m reflecting type Schmidt telescope, located in Xinglong Station in China and run by the Chinese Academy of Sciences. It has an effective aperture of 3.6--4.9 m and a wide Field-of-View (FoV) of 5{$^{\circ}$} \citep{2012Cuia}. There are 4000 optical fibers accommodated on the focal plane, each measuring 320 microns in diameter and covering 3.3$\arcsec$ in the sky. These fibers are fed into 16 low-resolution spectrographs and then registered on to a 4k{$\times$}4k CCD. The spectrum obtained for each object covers a wavelength range of 3650$-$9000 \AA, with a resolving power of $\sim$1800 around \textit{g} band \citep{2013Wang}. 
  
\begin{table}
\centering
\caption{List of steps followed to identify Classical Ae stars in the present study using LAMOST DR5.}
\begin{tabular}{cc} 
\hline
Criteria & Total candidates \\ 
\hline\hline
Total number of stars obtained from  & 439,920 \\
LAMOST DR5 (v3) A-type star catalog & \\
\hline
Stars selected based on the position & 4,787 \\
of H$\alpha$ emission line & \\
\hline
Selected stars after removing & 2,931 \\
multiple observations & \\
\hline
Stars having distance estimates in Gaia DR2 & 2,857 \\  
and photometric estimates in 2MASS \\
\hline
Stars obtained after removing spectra & 1,331\\
showing [S{\sc ii}], [N{\sc ii}], [O{\sc iii}] and [O{\sc i}] forbidden lines & \\ 
\hline
Stars selected based on the distribution & 1,282 \\
in 2MASS CCD & \\ 
\hline
A-type stars obtained by & 164\\
spectral template matching & \\
\hline
Ae stars identified excluding objects & 159 \\
showing high IR excess in SED & \\
\hline
\end{tabular}
\label{tab1:steps}
\end{table}

The five-year survey of LAMOST, classified as LAMOST Experiment for Galactic Understanding and Exploration (LEGUE) and LAMOST Extra GAlactic Survey (LEGAS), has released five sets of catalogs (DR1-5). The latest data release, LAMOST DR5, comprises objects from both the pilot survey and the following five surveys, accounting for the spectrum of 9,026,365 targets. Among this, 5,348,712 objects are classified as stars by the LAMOST pipeline. We retrieved the spectra of 439,920 stars from LAMOST DR5 A-star catalogue \citep{2019LuoVizieR}, which were used for this study (see Table \ref{tab1:steps}).

  
\section{Identification of Classical A\texorpdfstring{\MakeLowercase {e}}{e} stars}
\label{sec:Iden}

In this section, we explain the method for identifying CAe stars by analyzing their location in the near IR ($J-H$){$_0$} versus ($H-K${$_S$}){$_0$} diagram and in the optical CMD. In addition, we made use of all the available photometric data in optical and infrared regimes to estimate the IR excess emission in CAe stars from the spectral energy distribution. 

\subsection{Procedure for selecting the sample of Classical Ae stars}
\subsubsection{Selection of A-type stars with H$\alpha$ in emission}
\label{subsec: Halpha}

H{$\alpha$} emission is a prominent feature in different types of astrophysical objects such as galaxies, QSOs, CBe and Herbig Ae/Be stars (HAeBe). From the initial sample of 439,920 A-type stars, we identified those stars showing emission at H{$\alpha$}. Our H$\alpha$ detection is constrained in the wavelength window within 6561--6568 \AA. 
The H{$\alpha$} peak in these stars were identified using an automated python routine that employs \texttt{find{\textunderscore}peaks} function from the \texttt{scipy} package. The routine reports a peak if the flux at a sampling point is greater than the flux at adjacent points. In order to remove false detections due to noise peaks in the region, we made use of the in-built \texttt{width} parameter and only selected peaks that are at least 3 sampling points wide at the half-maximum. We also visually checked the resulting spectra for their peak selection quality. 

Using this method, we found 4,787 A-type stars that show H$\alpha$ in emission. However, it is noted that some sources have multiple observations from LAMOST. For these stars, we retained the spectrum with the highest SNR value in g band ($snr_g$). This reduced the number of unique Ae stars to 2,931. 
For these 2,931 stars, we queried the distance estimates from \cite{2018Bailer-Jones} and photometry from surveys such as 2MASS \citep{2003Cutri} and Gaia DR2 \citep{2018Gaia}, which returns a sample of 2,857 Ae stars. Stars that do not have distance estimates or photometric data in Gaia DR2 and 2MASS are eliminated from further analysis.

\subsubsection{Weeding out stars with forbidden emission lines}
  
While investigating the LAMOST spectra, we found that a considerable fraction of them show forbidden emission lines of [NII], [SII] and [OI] in their spectrum. B-type emission stars with forbidden lines in their spectra, known as B[e] stars, are considered to be distinct from the class of CBe stars. These forbidden lines are formed in an extended, rarer envelope, containing dust. Hence, B[e] stars show higher IR excess than CBe stars \citep{1976Allen, 2006Miroshnichenko, 2006Mirosh_review}. Moreover, \cite{1998Lamers} proposed that B[e] phenomenon is present among supergiants, PMS stars, etc., but not among main-sequence stars. 

We visually checked all the 2,857 Ae star spectra including their multi-epoch observations for forbidden lines, [SII] doublet 6717 {\AA}, 6731 {\AA}, [NII] doublet 6548 {\AA}, 6584 {\AA},  [OIII] 4363 {\AA}, 4959 {\AA}, 5007 {\AA} and [OI] 5577 {\AA}, 6300 {\AA}, 6363 {\AA}. From this sample, we found that 1,526 stars exhibit either one or all of the forbidden emission lines mentioned. Since the purpose of this work is to identify CAe stars, we do not further consider these Ae stars showing forbidden emission lines. Henceforth, we will be including a sample of 1,331 Ae stars for further analysis.
This sample of 1,331 stars we identified may belong to Classical Ae/Be (CAeBe) or HAeBe category. The H$\alpha$ emission-line is formed by recombination process in both these classes of stars. But, the nature and composition of the circumstellar disc giving rise to emission lines in these two groups are different. Hence, We used this IR excess to distinguish between CAe and CBe/HAeBe stars among our identified 1,331 stars.

\subsubsection{Separation of CBe and HAeBe stars using 2MASS color-color diagram}
\label{subsec:2mass}

In CBe/CAe stars, the IR excess is commonly attributed to thermal bremsstrahlung emission from free electrons in a hot, dense, ionized circumstellar disc \citep{1974Gehrz,1977Hartmann}. In case of PMS star discs, such as in HAeBe stars, dust is also present \citep{1993Gorti, 1998WatersWaelkens}. 
These stars show infrared excess in the continuum that suggest the presence of hot and/or cool dust in the circumstellar medium \citep{1992Hillenbrand}. Since dust emission is more intense than thermal bremsstrahlung, IR excess in HAeBe stars are relatively higher than CBe stars. \cite{1984Finkenzeller} suggested that the ($H$-$K_S$) colors of HAeBe stars are found to be greater than 0.4 mag and hence are separated from CBe or CAe stars in NIR color-color diagram (CCDm). 

\begin{figure}
\centering
\includegraphics[width=1\columnwidth]{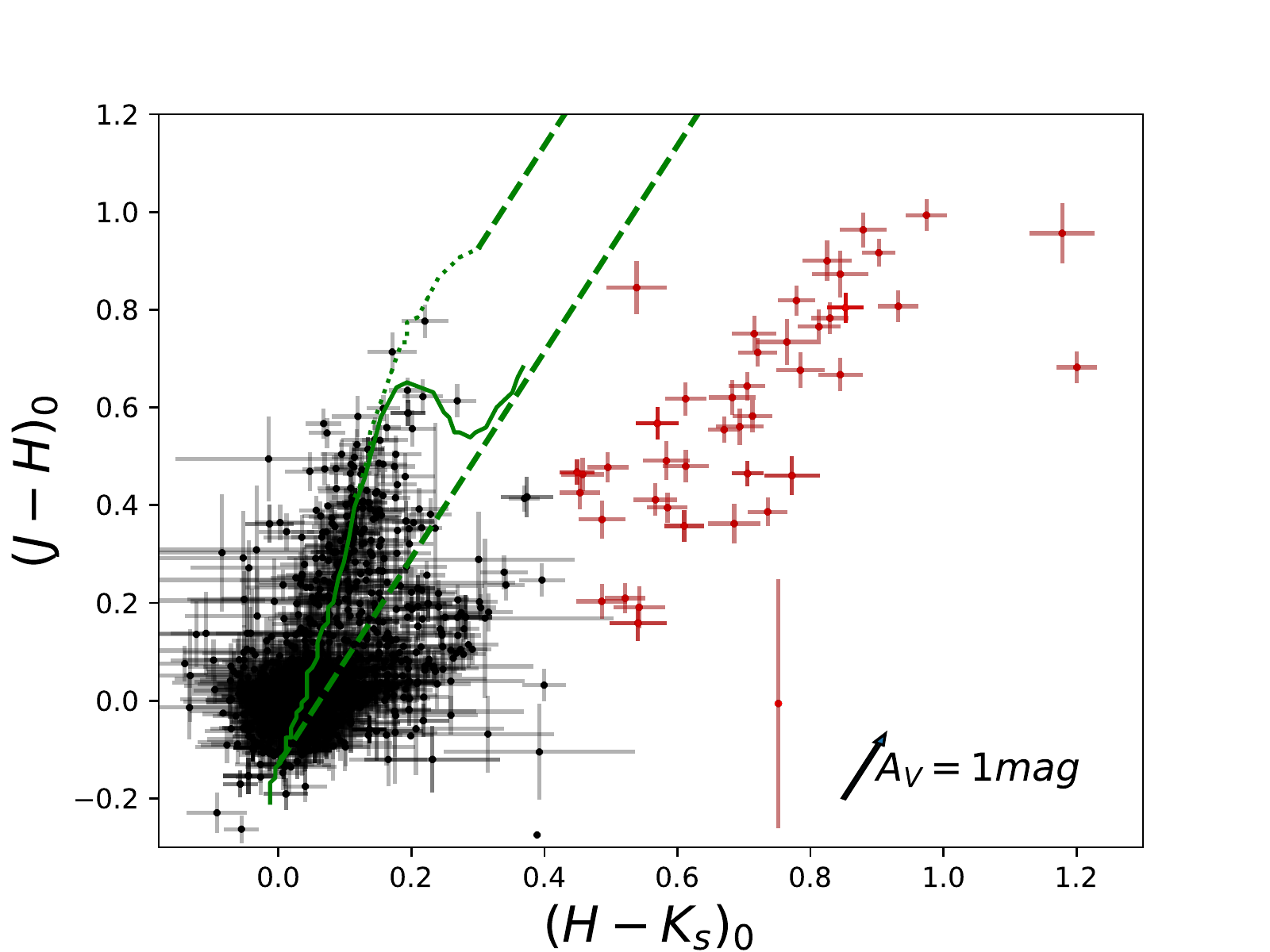}
\caption{2MASS $(J-H)_0$ vs $(H-K_S)_0$ color-color diagram of 1,331 Ae stars from LAMOST DR5, shown as black dots. The red dots represent 49 Ae stars identified as Herbig Ae/Be candidates, based on the criterion given by \protect\cite{1984Finkenzeller}. The error bars associated with $(J-H)_0$ and $(H-K_S)_0$ color is represented for each star. The solid, dotted and dashed lines represent the main-sequence, the giant branch and the reddening vectors, respectively, adopted from \protect\cite{1983Koorneef}, converted to 2MASS system using the relations from \protect\citet{2001Carpenter}}.
\label{2mass_ccd}
\end{figure}

We queried for the sample of 1,331 Ae stars in the 2MASS point source catalog \citep{2003Cutri} using VizieR for a search radius of 3$\arcsec$. The $J$, $H$, $K_S$ magnitudes obtained, needs to be corrected for extinction to represent them in the 2MASS CCDm. The extinction  parameter, $A_V$ values for all the stars are retrieved from the 3D dust maps of \cite{2019Green}, which were used for further analysis. 
The reddening corrected 2MASS CCDm of these Ae stars is shown in Fig. \ref{2mass_ccd}. Adopting the color criterion given by \cite{1984Finkenzeller}, we found 49 Ae stars are located to the right of reddening vector with $H-K_S$ $\geq$ 0.4 mag and hence can be considered as HAeBe candidates. However, since the present work is concerned with CAe stars, we removed this sample of HAeBe candidates. Henceforth, we consider 1,282 Ae stars for the analysis.

\subsubsection{Spectral type estimation}
\label{subsubsec:SpTyp}

\begin{figure*}
\begin{center}
\includegraphics[height=90mm, width=1.8\columnwidth]{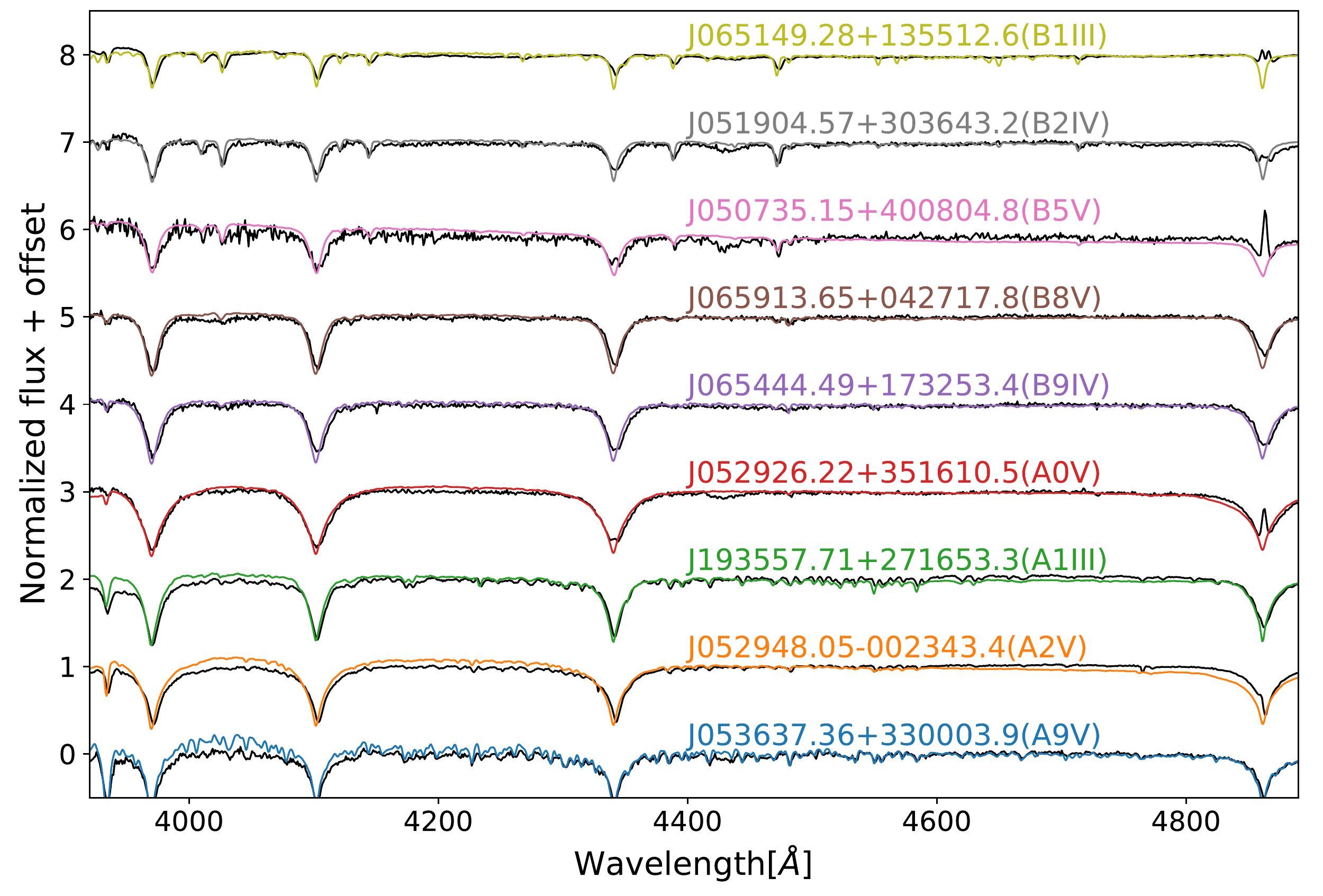}
\end{center}
\caption{The representative matching of the LAMOST Ae stars spectrum (black) and MILES spectral template (color) for different sub-types varying from B1 to B9, and then from A0 to A9, as we move from top to bottom. The spectra of 1,100 LAMOST Ae stars having SNR $>$ 10 are used for estimating the spectral type by matching templates. From this, we obtained a list of 164 Ae stars and 936 Be stars. The LAMOST ID of the star and the estimated spectral type by matching templates is mentioned above each spectrum. It is seen that the absorption strength of Ca{\sc ii} K and Balmer lines 3970 {\AA}, 4102 {\AA} and 4340 {\AA} increases from late B to A. Also, He{\sc i} lines are found to disappear as we shift towards A-type stars.}
\label{tempfit}
\end{figure*}   
Objects classified as stars by the LAMOST 1D pipeline in DR5, were divided into O- to M- spectral types based on five spectral line indices.
However, we found a mismatch up to 4 subtypes in some of the B and A-type stars, when compared with the spectral type from previous studies. The spectral match between LAMOST estimates and literature (mostly from SIMBAD) gets better for A-type stars (48\%) when compared to O- and B-type stars ($\sim$22\%). 
The spectral type of stars is estimated by the LAMOST instrument team based on a spectral template matching technique. Spectral templates were constructed using data from the pilot survey and LAMOST DR1 \citep{2014Wei}. A set of 183 stellar spectra serve as templates in the pipeline for classifying stars to each spectral type with a high correlation ratio \citep{2019Kong}. Though, in A spectral type, there exist more than one template for each subclass, O- and B- types have only four templates due to the lack of OB- stars in LAMOST DR1 \citep{2014Wei}. 
So, it is quite possible that the stars which were catalogued as early A-type in LAMOST can equally be late B-type. 
  
Hence, we re-estimated the spectral type for stars present in the LAMOST A-star catalog. But, before we proceed, it has to be noted that the low signal-to-noise ratio (SNR) in the spectrum can affect the spectral type estimation. In the LAMOST spectrum of each object, SNR is included in the header of the concerned FITS files for different bands such as \textit{u, g, r, i} and \textit{z}. The data quality for LAMOST targets are considered as good where the SNR $>$ 10 in \textit{g} or \textit{i} bands \citep{2015Luo}. We checked the SNR information for different bands, given in the header of each spectrum used for estimating the spectral type. 

It is worth noting that, ionized helium (He{\sc ii}) lines are present in O-type spectrum whereas B-type stars show neutral helium (He{\sc i}) lines, which will essentially disappear in the spectra of A-type stars \citep{2009Gray}. The Balmer series of hydrogen lines are the prominent absorption features in the spectrum of A-type stars. 
We observed in many cases, spectra classified as A-type show visible He{\sc i} lines. Similar discrepancies in spectral type estimation using LAMOST have also been observed by \cite{2015Liu}, \cite{2015bZhong} and \cite{2019Liu}. Hence, we used the MILES stellar library \citep{2006MILES} as templates for spectral type estimation because the resolution is similar to that of LAMOST . It consists of spectra of $\sim$980 stars, covering a wavelength range of 3525--7500 \AA, at a spectral resolution of 2.5 \AA~ \citep{2011MILES2}. All the spectra present in the stellar library were observed with Isaac Newton Telescope (INT), located at La Palma in the Canary Islands, Spain. This ensures homogeneity among the data. The template stars cover a wide parametric space, i.e. T{$_{eff}$} $=$ 3000 -- 40,000 K, log ({$g$}) = 0.2 -- 5.5, when compared to other spectral libraries  available in the literature. We found that 79 library spectra are available in the spectral range B0 -- A9 covering luminosity classes {\sc III} -- {\sc V}.

To receive a better control of the spectral type, we visually identified the spectra where lines used for classification are not contaminated by noise. After a thorough scrutiny, we obtained 1,100 Ae spectra with good SNR for spectral classification. For these selected sample of stars, we fitted the LAMOST spectra with B0--A9 templates from the MILES spectral library. MILES spectra that fit the best visually were used to define the spectral type of CAe stars, which are listed in Table \ref{Table2:final_CAe}.
For A-type stars, we used the absorption strength of Mg{\sc ii} 4481 \AA~, Ca{\sc ii} 3933 {\AA} and Balmer series of lines from H$\epsilon$ to H$\delta$ for spectral type estimation. While performing spectral classification of these stars, we also checked for the absence of He{\sc i} and He{\sc ii} absorption lines.
Similarly, for B-type stars, we used the absorption strength of He{\sc i} 4009 \AA, 4026 \AA, 4144 \AA, 4388 \AA~and 4471 \AA~lines for spectral type estimation. We did not use H$\alpha$, H$\beta$ and H$\gamma$ for the analysis since they are present either in emission or show an emission component in the absorption profile. 

After carefully matching the spectra, we obtained a list of 164 Ae stars with spectral types from A0$-$A6 and 936 Be stars with spectral types from B1$-$B9. 
A representative example of this spectral type estimation using template match is shown in Fig. \ref{tempfit}. Though the spectral type of the Ae stars are quiet similar to the estimates made by the LAMOST 1D pipeline (deviates only by one or two sub-types), it has to be noted that for Be stars a deviation of about 5 sub-classes are seen. The detailed analysis of this sample of CBe stars is reserved for another study. Further, in this work, we will be concentrating on the analysis of 164 CAe stars. The sample of 164 Ae stars may also contain PMS Herbig stars. So in the following sub-section, we extended our analysis on to the IR excess by constructing the spectral energy distribution of the Ae stars using available optical/infrared photometry. 



\subsubsection{Spectral Energy distribution of the Classical Ae stars}
\label{subsec:SED}
 
\begin{figure*}
\includegraphics[height=105mm, width=175mm]{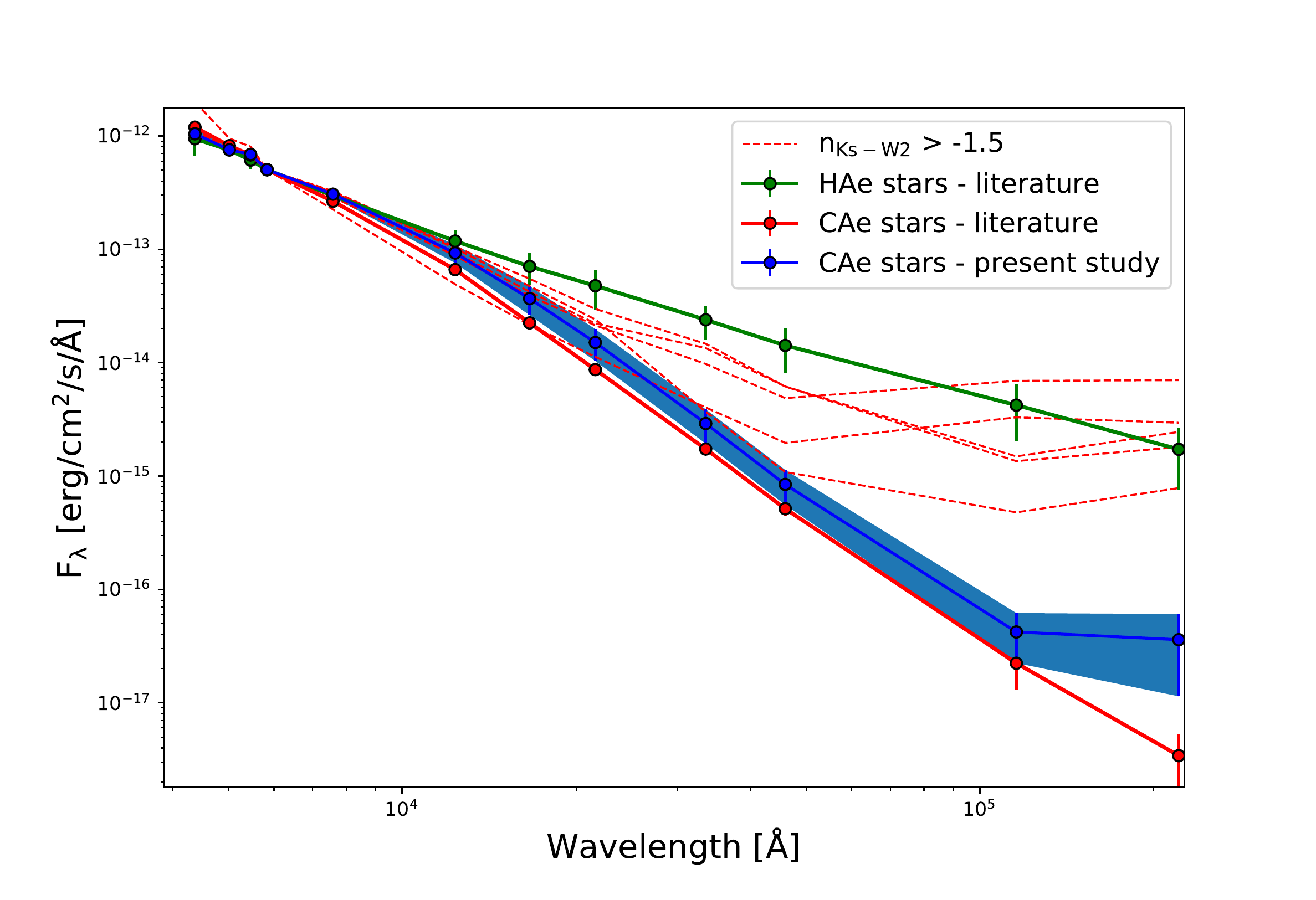}
\caption{The spectral energy distribution of 131 CAe candidate stars with available photometric magnitudes in optical and near-infrared bands, identified from the present study is shown as the blue shaded region. The 5 Ae stars that belong to HAe region is denoted in terms of n{$_{K{_s} - W2}$} > -1.5, indicated as red dashed lines. The green and red model lines represent the median SED grids for Herbig Ae and Classical Ae stars from literature, respectively. The maximum and minimum divergence of each of these distributions are represented as the absolute deviation in error bars. The high IR excess and the Lada index, n{$_{K{_s} - W2}$}, of the the stars represented in red dashed lines suggest that they can be Herbig Ae candidates.}
\label{Fig:SED}
\end{figure*}

In order to construct the spectral energy distribution (SED), we queried for the photometric magnitudes of the 164 Ae stars by cross-matching object positions with Gaia DR2 \citep{2018Gaia}, APASS DR9 \citep[][for optical]{2015APASS}, 2MASS All-Sky Catalog \citep[][for near-IR]{2003Cutri} and WISE All-Sky Data Release \citep[][for mid-IR]{2012WISE}. Out of 164 Ae stars, 131 have photometric magnitudes available in optical and near-infrared bands. The observed magnitudes of CAe stars in different photometric bands are then corrected for corresponding interstellar extinction, adopting conversion relations from \citet{1990Mathis}. These extinction corrected magnitudes are used to construct the SED for our target stars. 
The SED of CAe stars identified from this study is shown in Fig. \ref{Fig:SED}. In order to construct the median grids, a sample of 50 HAe stars are taken from \cite{2019Arun} including only stars with spectral types between A0-A9. On the other hand, the median of CAe stars is constructed using the 21 CAe stars identified from the literature.
From the SEDs, we notice that 5 stars out of 164 are located close to the SED grid of HAe stars. 


In the case of young stellar objects (a broader category encompassing PMS and protostars), IR excess is attributed to the presence of circumstellar dust, which absorbs visible light from the young star and re-radiates in the infrared. The closer the dust is to the star, the hotter it is and the shorter is the characteristic wavelength of its infrared emission. As the distribution of dust moves away from the star, it radiates at longer wavelengths as cool dust \citep{1992Hillenbrand, 1999Malfait}. This emission appears as excess flux over the stellar continuum in the infrared region of the SED. 
The presence of hot/cool dust in the disc of such stars can be determined from the slope of the IR region in the SED. \cite{1987Lada} quantified a classification scheme for YSOs depending on their slope in the IR region of the SED, known as Lada indices. Based on the steepness of the indices at various wavelength intervals, objects are classified into different categories such as Class 0, Class I, Class II and Class III \citep{1993Andre}. We estimated the n{$_{2-4.6}$} Lada index of all the 164 Ae stars using 2MASS \textit{K\textsubscript{S}}  and WISE \textit{$W2$} magnitudes, which is similar to the index calculated by \cite{2019Arun} for HAeBe stars. 
We find that the 5 stars having SED closer to that of HAe star grids also show index n{$_{2-4.6}$} $>$ -1.5, indicating them to be HAe stars.

Inspecting the literature, we noticed that from these 5 stars, 3 stars have been identified as PMS candidates \citep{2010Beerer, 2014Rapson, 2018Cottle}. In order to re-assure our classification, we examined their {$V$}-[12] magnitude color excess. CBe stars and HAeBe stars are well separated on a {$V$}-[12] color excess versus temperature diagram \citep{1992Hillenbrand}, due to the difference in the thermal emission from the disc. The IR excess in HAeBe stars is considerably larger than that of CAeBe stars at a given temperature. For example, at 10,000 K, CAe stars show $V$-[12] color excess close to zero, while it is from 3--10 mag for HAe stars \citep{2003Monin}. 

To obtain the magnitudes of CAe stars in IRAS 12 {$\mu$}m, we searched in IRAS Point Source catalogue. Since most of our targets are not detected in IRAS, we used the magnitudes in WISE $W3$ \citep{2012WISE} as an equivalent substitute. This is combined with the V magnitude values obtained from the APASS survey \citep{2015APASS} to estimate the respective {$V$}-{$W3$} values for our sample CAe stars. This color is corrected for reddening using the $A_V$ values estimated in this work. The mean wavelength of \textit{N}-band filter falls at 10.3{{$\mu$}m} and the range extends from a minimum of 7.0{{$\mu$}m} to a maximum of 13.6{{$\mu$}m}. Taking advantage of this wide range, we used (V-N){$_0$} values from \cite{1966Johnson} for intrinsic (V-[12]){$_0$} color. Further, we estimated the (V-W3) color excess of all the 5 stars, which is found to be $>$ -1.5 mag, suggesting them to be HAe stars and hence excluded from further analysis. 

\subsubsection{Classical Ae stars in the optical color-magnitude diagram (CMD)}
\label{subsec:CMD}

\begin{figure}
\includegraphics[width=1\columnwidth]{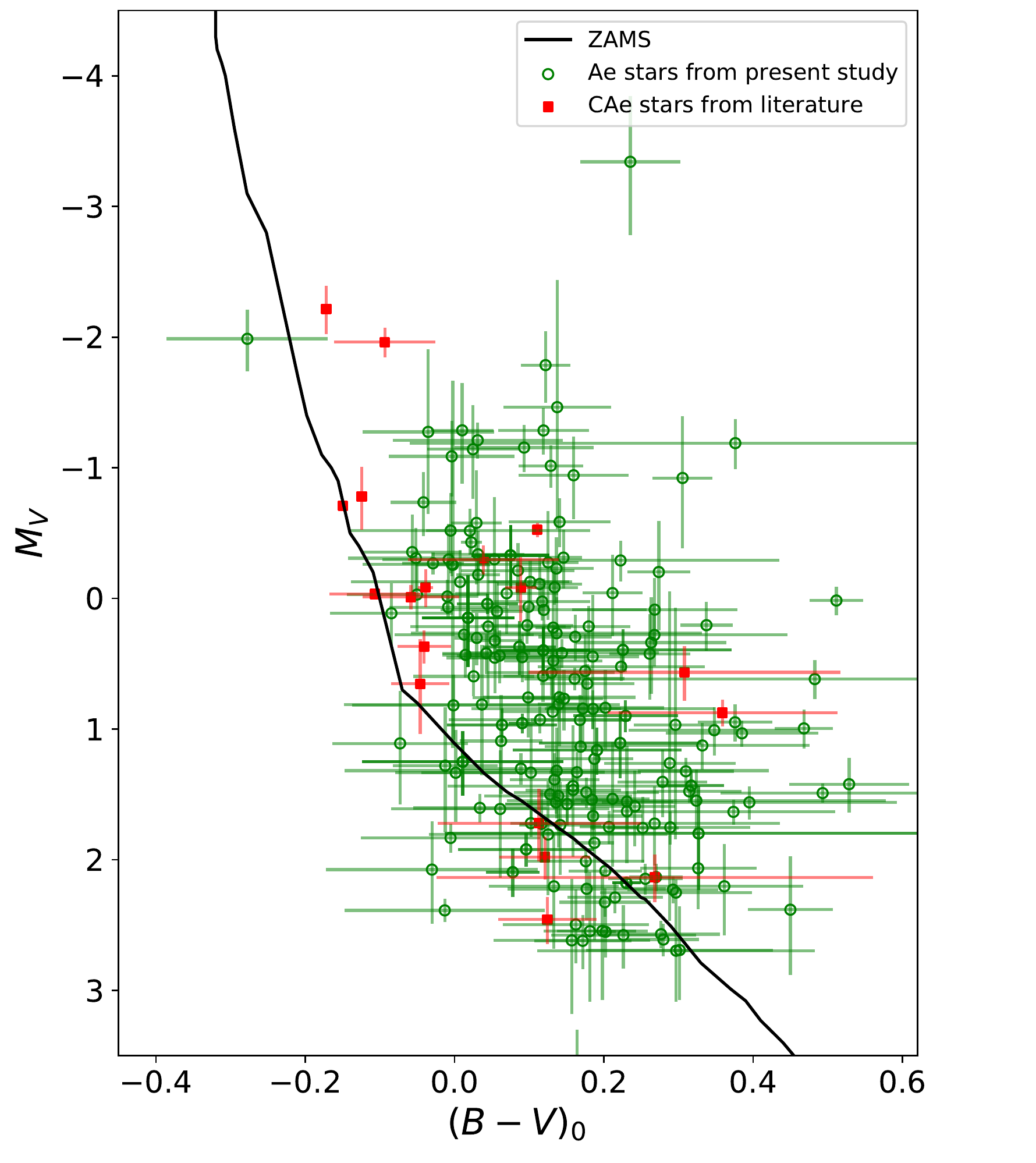}
\caption{Optical color-magnitude diagram of 126 CAe stars from our study. The black line represents ZAMS from \protect\cite{2013Pecaut}. Ae stars in our analysis is shown in open green circles. The M{$_V$} and (B-V){$_0$} error for each star is also represented. Classical Ae stars obtained from the literature is marked in filled red squares. We observe that the Ae stars are distributed close to the main-sequence, sharing the same region as that of CAe stars from literature.}
\label{opticalcmd}
\end{figure}

In order to inspect the position of our identified CAe stars in the optical CMD, the optical magnitudes in \textit{B} and \textit{V} filters is queried from the APASS survey \citep{2015APASS}. These magnitudes are obtained for 126 CAe stars, extinction corrected and converted to absolute scale using the Gaia DR2 distance estimation from \citet{2018Bailer-Jones}. The position of these 126 stars in the optical CMD is compared with 18 CAe stars compiled from the literature, which has optical magnitudes available from the APASS survey. The extinction corrected CMD, M$_V$ vs $(B-V)_0$, is shown in Fig. \ref{opticalcmd}. The higher error in $(B-V)_0$ is due to the error associated with the $B$ and $V$ magnitudes in the APASS survey. From the figure, we observe that our sample of CAe stars are present close to the ZAMS, indicative of them being main-sequence stars. It is also noted that these CAe stars share the same region in CMD as that of CAe stars obtained from the literature, which makes our classification much certain.

\begin{figure}
\includegraphics[width=1\columnwidth]{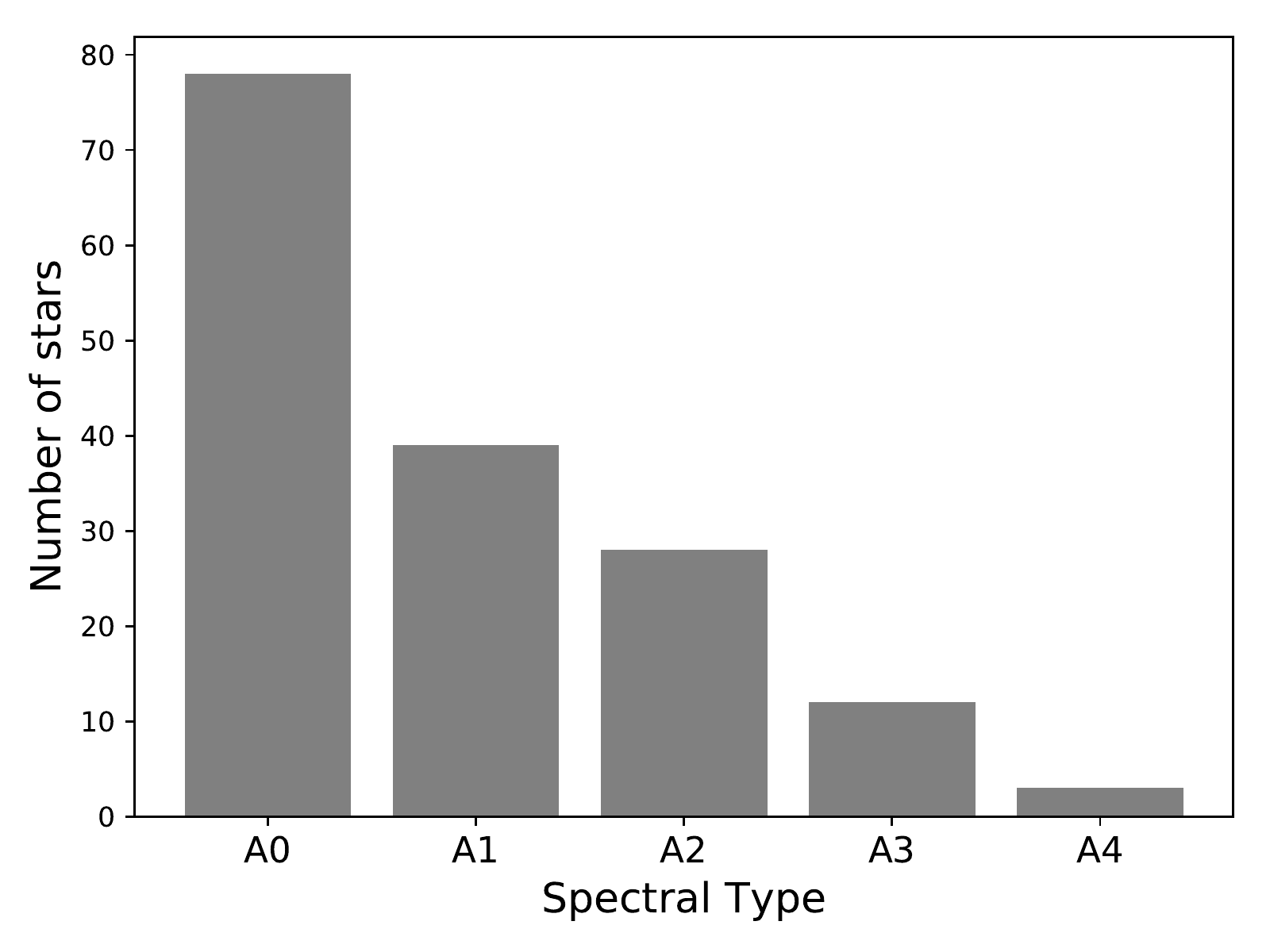}
\caption{Histogram showing the incidence of 159 CAe stars from this study. It is observed that the distribution of CAe stars in our sample also peaks at A0 and steadily decreases towards later spectral types, as suggested in previous studies.}
\label{SpTyp_num}
\end{figure}

\begin{figure*}
\includegraphics[width=2.15\columnwidth]{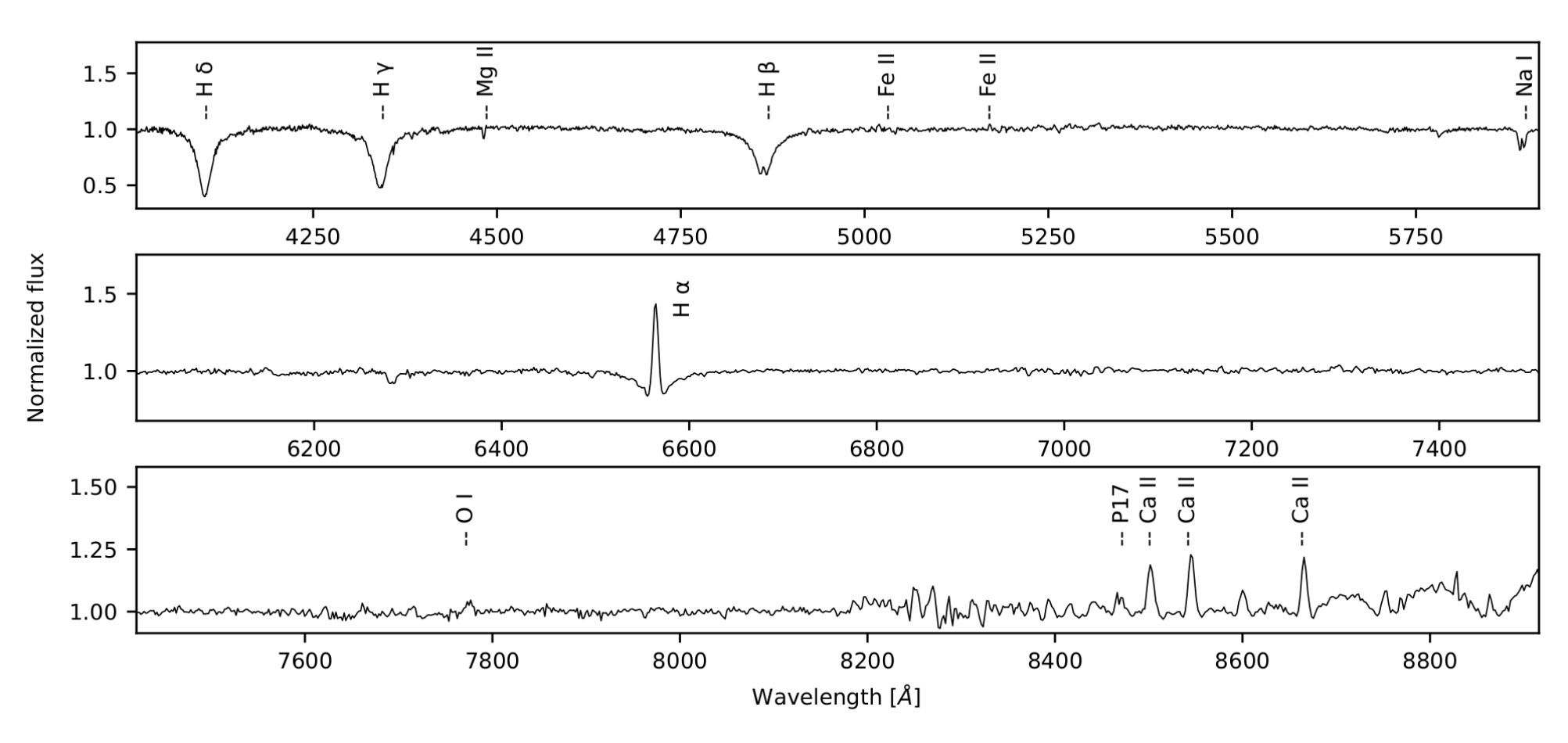}
\caption{Representative spectrum of a CAe star from the present study, named J200257.09+453007.9, having spectral type A0V. Prominent spectral features are labelled.}
\label{repplot}
\end{figure*}

  
      
\section{Results and Discussion}
\label{sec:result}

\subsection{New Ae stars identified using LAMOST DR5}
\label{subsec:NewAe}

After applying tight constraints to a sample of 2,931 Ae stars from the LAMOST survey, we identified 159 new CAe stars through our analysis. The near and mid-IR excess of these stars are similar to that of CBe stars. From our sample, we observed that the number of CAe stars are peaking at A0 spectral type and gradually decrease towards late-types, as suggested by \citet{1986Andrillat}. The histogram of spectral type distribution
for our sample of CAe stars is shown in Fig. \ref{SpTyp_num}. Our study has identified CAe stars beyond A3 spectral type for the first time. In addition to the CAe stars identified by previous literature, our new detection of 159 CAe stars increase the total number of known CAe stars to 180. The list of CAe stars identified from this study is presented in Table. \ref{Table2:final_CAe} showing the spectral features we observed for them. A representative spectrum of one of our identified CAe star, named J200257.09+453007.9 is shown in Fig. \ref{repplot}.
 

\subsection{Observed spectral features in our sample}
\label{subsec:specfea}
In this section, we discuss about all major emission lines observed in our sample of CAe stars, covering the whole wavelength range of 3700--9000 {\AA}. Among our sample, 45 stars have multi-epoch observations in LAMOST. The prominent features in each spectra is mentioned in Table \ref{Table2:final_CAe} for 15 of our sample stars. We also provide a comparative study of the optical spectra of our newly identified CAe stars with that of well studied CBe stars.

\subsubsection{Balmer series of Hydrogen lines}
\label{Balmer}
In CBe stars, recombination emission lines belonging to Balmer series originate from the circumstellar disc, H{$\alpha$} being the most prominent among them \citep{2005Tycner}.
It is necessary to study these lines as they provide insights about the circumstellar environment and interaction of photospheric radiation with the disc \citep{2006Grundstrom}. 
\citet{Gourav} observed that the H{$\alpha$} EW values for CBe stars are mostly lower than -40 {\AA}. \cite{2013Rivinius} suggested that emission-line A-type stars are expected to show weak H{$\alpha$} emission compared to CBe stars. In this context, we checked the H$\alpha$ EW distribution for our sample stars. 

\begin{figure}
\includegraphics[height=118mm, width=1\columnwidth]{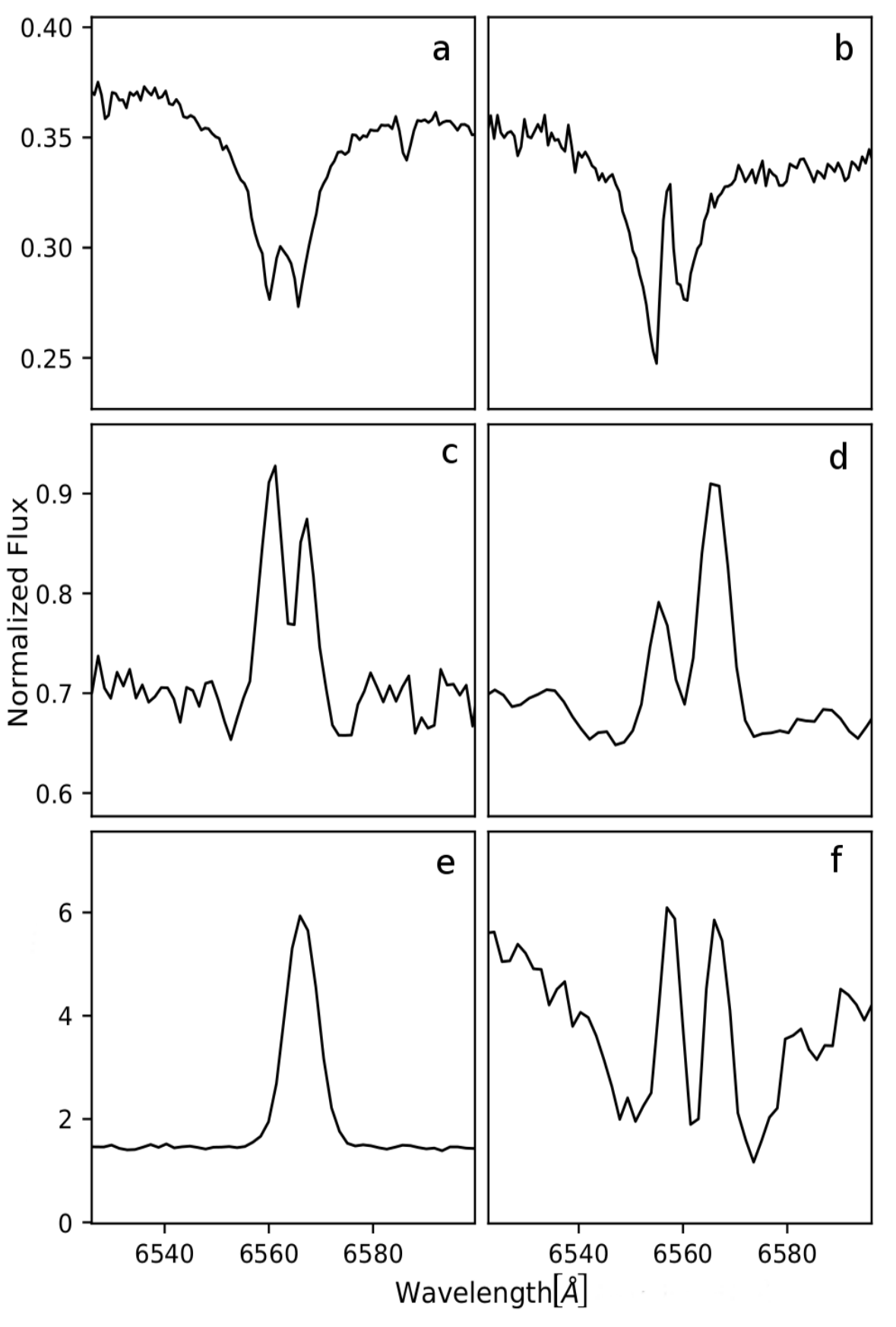}
\caption{H{$\alpha$} emission-line profiles observed in our sample of 6 CAe stars. (a) J051256.73+472841.3 and (b) J062812.14+185018.6 shows emission in absorption (\textit{eia}) profiles, (c) J005911.26+555720.1 and (d) J001742.37+454454.4 shows double-peak emission (\textit{dpe}) profiles, (e) J064714.86+035830.0 shows single peak emission (\textit{e}) profile and (f) J011650.55+492723.3 shows double-peak emission in absorption profile.}
\label{Halpha_profiles}
\end{figure}

Out of 159 CAe stars, 139 (87\%) show H$\alpha$ emission in absorption ({\it eia}) profile, including the weak emission-line stars. Such a profile is generally less observed in case of CBe stars. We found that 16 stars show double-peak H{$\alpha$} emission ({\it dpe}), while single peak emission is observed only in 4 cases. Fig. \ref{Halpha_profiles} presents a representative sample of the H{$\alpha$} line profiles observed in the spectra CAe stars from this study.
We estimated the H$\alpha$ emission EW using standard \texttt{IRAF} tasks.
The measurement error in our EW estimates is {$\sim$}10\%. In emission-line stars, H$\alpha$ appears in emission only after filling the absorption profile generated from the photosphere of the underlying star. Hence, the absorption EW should be added to the emission EW to calculate the effective H$\alpha$ emission EW for a CAe star. We measured the absorption strength at H{$\alpha$} from the synthetic spectra using models of stellar atmospheres \citet{1993Kurucz} corresponding to the spectral type of each star. This underlying photospheric contribution is added to the measured emission EW to estimate the corrected H{$\alpha$} EW.

\begin{figure}
\includegraphics[width=1\columnwidth]{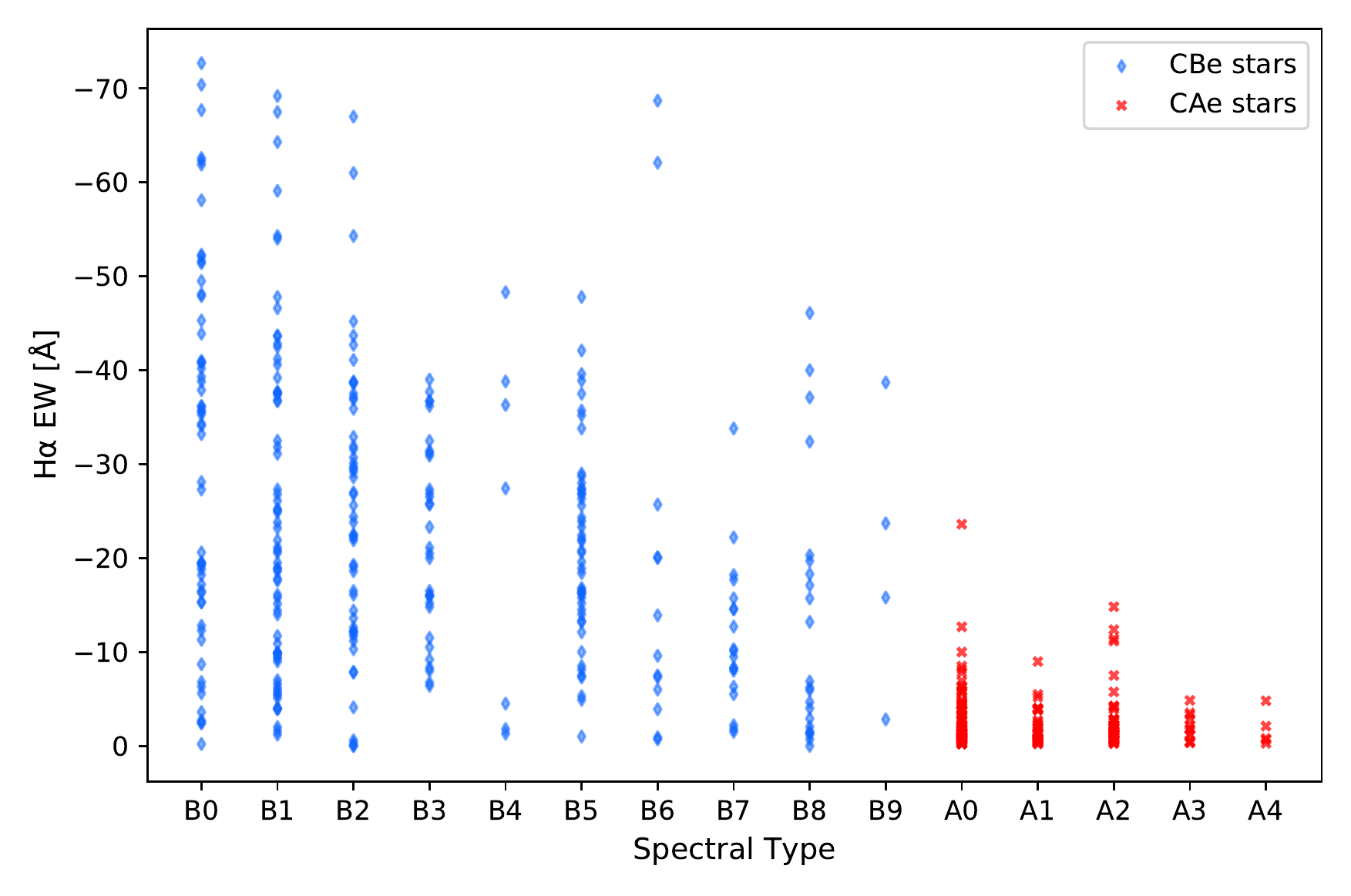}
\caption{The distribution of H$\alpha$ emission EW for our sample of CAe stars (red) shown along with CBe stars (blue). The H{$\alpha$} EW of 323 CBe stars were adopted from \protect\cite{2011Mathew}, \protect\cite{2013Raddi}, \protect\cite{2013Barnsley}, \protect\cite{2013Catanzaro} and \citet{Gourav}. We notice that CAe stars show comparatively weaker emission and there exist a steady decrease in emission strength from CBe to CAe stars.}
\label{halpha_sptyp}
\end{figure}

The corrected H{$\alpha$} EW ($H{\alpha}EW_c$) for our sample CAe stars are within the range of -0.2 to -23.6 \AA, which is comparatively lower than that reported for CBe stars. The $H{\alpha}EW_c$ of our sample of CAe stars is listed in Table \ref{Table2:final_CAe}. The distribution of {$H{\alpha}EW_c$} against each spectral type is shown in Fig. \ref{halpha_sptyp}.     
Assessing the {$H{\alpha}EW_c$} distribution from B0 to A4, we see a steady decrease in the emission strength. However, it may be noted that there is uneven peaking in emission strength for some sub-types. This result may be suggestive of the fact that the disc size in CAe stars might be considerably less than that of CBe stars since the H$\alpha$ emission strength is correlated with the extent of the disc from which it is formed.

  
H$\beta$ is present in absorption in 151 (94\%) of our CAe stars, whereas the rest 8 show {\it eia} profile. H$\beta$ EW for our sample ranges between -0.3 to -11.2 {\AA}, similar to what is seen in previous studies for CBe stars \citep[e.g.][]{2011Mathew, 2012Paul}. The H{$\gamma$} and H{$\delta$} profiles for all our CAe stars are visible in absorption. It is interesting to notice that although the H$\alpha$ EW range of CAe stars are lower than CBe stars, H$\beta$ EW range is comparable. 
  
\subsubsection{Paschen series of Hydrogen lines}
In CBe stars, Paschen emission lines starting from P9 or P10 to P23 are visible in the wavelength region 7500--10000 {\AA} \citep[e.g.][]{1981Briot, 2011Mathew, 2012Mathew}. \cite{1967Andrillat} noticed Paschen emission lines till P26 for their sample of CBe stars. But as H{$\alpha$} emission strength diminishes from spectral type B towards A, one expects Paschen emission strength to also diminish \citep{1990Andrillat}. \cite{1981Briot} noticed that Paschen emission lines in CBe stars are usually observed in early spectral types. 

In our sample, we noticed absorption lines of Paschen lines from P11 to P17 in 154 cases, while only 3 cases show in emission.
We notice a positive correlation between Paschen emission strength and H-alpha emission strength but the sample is too small (3) to make a statistical correlation analysis.



\subsubsection{Fe{\sc ii} emission lines}
Fe{\sc ii} emission lines are commonly present in the spectra of CBe stars \cite[e.g.][]{1987Hanuschik, 1992Slettebak, 2011Mathew}. Among different Fe{\sc ii} emission lines (belonging to various multiplet series) found in CBe stars, Fe{\sc ii} 5169 {\AA} (multiplet no. 42) line appears to be the strongest in optical regime \citep{1987Hanuschik}. \citet{Gourav} observed that in CBe stars, Fe{\sc ii} 5169 \AA~emission-line EW peaks at B1--B2 spectral types. Moreover, it has been suggested by various authors that Fe{\sc ii} emission strength fades after B5 in CBe stars \cite[e.g.][]{1987Jaschek, 1988Andrillat, 1995Ballereau}.

Hence, we do not expect Fe{\sc ii} emission lines in CAe stars since Fe{\sc ii} emission strength steadily decreases and fades out by late B-types. But interestingly, we found 8 stars exhibiting Fe{\sc ii} 5169 {\AA} emission line, whereas Fe{\sc ii} 5317 {\AA} is the second most commonly visible line (5 stars). Apart from these, Fe{\sc ii} 7712 {\AA} (multiplet no. 73) is observed in 2 cases and Fe{\sc ii} 5363 {\AA} (multiplet no. 48), 5198 {\AA}, 5235 {\AA} and 5276 {\AA} (multiplet no. 49) lines are visible in emission in one star each. Hence, it is quite surprising to notice Fe{\sc ii} emission lines in CAe stars. We suggest further comparative analysis to be done including CBe stars to understand whether the emission strength fades towards late B-types and starts building in A-types due to some mechanism.  

\subsubsection{O{\sc i} emission lines}
O{\sc i} 8446 \AA~is another commonly observed emission feature in CBe stars, mostly in emission but occasionally in absorption also \citep{2011Mathew, Gourav}. This line gets blended with the P18 line whenever both are present in emission \citep{Mathew_2012b}. O{\sc i} 7772 \AA, the counterpart of O{\sc i} 8446 {\AA}, is also visible in CBe stars in some cases. In CBe stars, it is mostly seen in stars with spectral types earlier than B3 \citep{1988Andrillat, Gourav}.  

Only 6 stars of our sample of 159 show O{\sc i} 8446 \AA~in emission. O{\sc i} 7772 {\AA}, the counterpart of O{\sc i} 8446 {\AA}, is also observed in CBe stars with spectral type earlier than B3. On the contrary, we observed O{\sc i} 7772 {\AA} in 98 CAe stars, out of which 14 stars show it in emission, whereas it is in absorption in all other 84 cases.


\subsubsection{Ca{\sc ii} triplet lines}
Ca{\sc ii} triplet (8498{\AA}, 8542{\AA} and 8662{\AA}) emission lines are also observed rarely in CBe stars \cite[e.g.][]{1976Polidan, 1988Andrillat, 2011Mathew, 2018Shokry, 2019Klement}. The formation mechanism and region where these lines are produced in the disc of CBe stars is still poorly understood. Recently, \citet{Gourav} suggested that in CBe stars, Ca{\sc ii} triplet emission can originate either in the circumbinary disc or from the cooler outer regions of the envelope, which might not be isothermal in nature. \cite{1990Andrillat} suggested that Ca{\sc ii} emission lines are rarer in stars with spectral type later than B6. 

Ca{\sc ii} triplet are low ionization lines which require a region having cooler temperature (T$\sim$~5000 K) to form \citep{1976Polidan}. Hence, one can expect more number of Ae stars to show Ca{\sc ii} lines in emission due to their lower temperature compared to CBe stars \citep{1995bcesbkJaschek}. But surprisingly, we noticed that Ca{\sc ii} triplet is in emission in 16 CAe stars ({$\sim$} 10\%), which is lower when compared to the fraction ({$\sim$} 15--27\%) of CBe stars showing these lines in emission (\cite{1976Polidan, 1988Andrillat, 2018Shokry, Gourav}. 


To summarize, emission lines of O{\sc i}, Fe{\sc ii} and Paschen series are not expected to be seen in A-type stars. However, we observe them in some of our newly discovered sample of 159 CAe stars. Similarly, more CAe stars should have shown Ca{\sc ii} emission lines, if these lines are formed in the circumstellar disc. But, we found the fraction of stars showing Ca{\sc ii} triplet emission to be lower than CBe stars. More importantly, the H{$\alpha$} emission strength shows a steady decrease from late-B to Ae stars. These aspects can be understood in future through much detailed further analysis of these CAe stars.

\section{Conclusions}
\label{sec:concl}
In this paper, we studied a sample of 2,931 A-type stars showing H$\alpha$ in emission, obtained from LAMOST DR5 and report the detection of 159 CAe stars through spectral analysis, all of which can be considered as new detection. Our primary motivation was to identify CAe stars from this sample, which is a rare class of objects that demands attention. This is the first systematic study performed till date to identify and characterize a sample of CAe stars from any database of emission-line stars. The important results from this study is summarised below.
\begin{itemize}
\item Making the best use of the available multi-wavelength photometric and astrometric information and through the analysis of spectral features, we identified a sample of 159 CAe stars, improving the known number of CAe stars ({$\approx$}21) about nine times. We found that their spectral type ranges between A0 -- A4. Our analysis identified CAe stars beyond A3 spectral type for the first time. From our sample, we also observed that the frequency of CAe stars is maximum at A0 and gradually decreases towards late types, as suggested by \citet{1986Andrillat}.

    
\item The emission line stars used as the starting sample in this study was catalogued as A-type stars using the automated spectral type estimation algorithm by the LAMOST team. However, we found mismatch between this spectral type and those listed in the literature for some stars, particularly for B-type and early A-type stars. Hence, we re-estimated the spectral type of Ae stars in our sample using the template matching technique with the standard spectral templates from MILES library \citep{2006MILES}. We found that only 32 stars from our sample have spectral type available from literature. By re-estimation of spectral type of Ae stars from LAMOST DR5, we found that in 26 cases, our newly estimated spectral types match with those in literature. Some of the stars mentioned as Ae in the LAMOST catalogue are detected to be B-type stars from our classification. This is further substantiated due to the presence of He{\sc i} absorption lines in the spectrum, which is generally seen in stars earlier than B5.  

\item Our study is also the first spectroscopic survey for a sample of CAe stars, thus providing an atlas of emission lines present in CAe stars for the first time. H$\alpha$ is visible as {\it eia} profile for 87\% of our identified CAe stars. The H$\alpha$ EW range for our sample ranges between -0.2 to -23.6 \AA, which is comparatively lower than that in CBe stars. The emission-line A-type stars show weak H{$\alpha$} emission compared to CBe stars as suggested by \cite{2013Rivinius} and a steady decrease in emission strength from late-B to Ae stars is noticed in our sample. We suggest for further studies to confirm if the disc size in CAe stars are be considerably lesser than that of CBe stars, as it correlates with H{$\alpha$} emission strength.


     
\item Interestingly, we also found emission lines of Fe{\sc ii}, O{\sc i} and Paschen series in some of our sample stars. This is a surprising result since previous studies suggest that the emission strength of these lines reduce towards late B-types and disappear altogether for A-type stars.
    
\item Ca{\sc ii} triplet emission lines are observed in 16 (10\%) of our sample CAe stars, which is lower when compared to the fraction ({$\sim$} 15--27\%) of CBe stars showing these lines. This result shows the need for further investigations to understand the physical conditions required in the disc of CBe and CAe stars for the formation of Ca{\sc ii} emission lines.  
    \end{itemize}

Our study is to motivate the Be star community about the need to further detect and study CAe stars. Since the `Be pheomenon' is still poorly understood even after 150 years of CBe star research, detection of more CAe stars can provide new insights about various disc formation mechanisms in CAe and also CBe stars. We could not perform any analysis about the rotation velocity distribution of CAe stars due to the low resolution of LAMOST spectra. We plan to perform such an analysis for our newly identified sample of CAe stars using high resolution spectra in future.



\section*{Acknowledgements}
We would like to thank our referee, Dr. David Bohlender for providing helpful comments and suggestions that improved the paper. We also like to thank our colleague, Mr. Ujjwal Krishnan for providing us technical support in improving our work.
B.M. would like to thank the Science \& Engineering Research Board (SERB), a statutory body of Department of Science \& Technology (DST), Government of India, for funding our research under grant number CRG/2019/005380.
We would like to thank the support given by Center for research, CHRIST (Deemed to be University), Bangalore, India. This work has made use of data products from the Guo Shoujing Telescope (the Large Sky Area Multi-Object  Fibre  Spectroscopic  Telescope,  LAMOST). This work has used the Gaia DR2 data to re-estimate the distance and extinction parameters for stars present in our analysis. Hence, we express our gratitude to the Gaia collaboration for providing the data. We also thank the SIMBAD database and the online VizieR library service for helping us in literature survey and obtaining relevant data.

\section*{Data Availability}
The data underlying this article were accessed from the Large sky Area Multi-Object fibre Spectroscopic Telescope (LAMOST) data release 5 (http://dr5.lamost.org/). The derived data generated in this research will be shared on a reasonable request to the corresponding author.

\bibliographystyle{mnras}
\bibliography{example}

\begin{landscape}
\centering
\begin{table}
\caption{List of the Classical Ae stars identified from this study. The extinction values retrieved from \citet{2019Green}, Spectral Type obtained from template matching with MILES library spectra{$^{(1)}$}, Spectral classification given by LAMOST{$^{(2)}$}, $B$ and $V$ magnitudes from \citet{2015APASS}, Lada index (n{$_{(2-4.6)}$}) calculated from the spectral energy distribution, observation date of the star in the LAMOST survey, the measured H{$\alpha$} EW corrected for underlying photospheric absorption and the prominent spectral features found in the spectra from this study are provided. Full version of this table is made available online.}
\label{Table2:final_CAe}
\begin{tabular}{lcccccccccl}
\hline
LAMOST ID & A$_V$ & (H-K)$_0$ & Spectral & Spectral & B & V & n{$_{(K_s-W2)}$} & Observation & H{$\alpha$}EW{$_{\textit{c}}$} & Other prominent\\
 & mag & mag & Type$^1$ & Type$^2$ & mag & mag &  & Date & {\AA}& spectral features\\
\hline
\hline
J001742.37+454454.4	&	0.2	&	0.13	&	A3V	&	A6V	&	13.24	&	12.83	&	-2.77	&	2015-10-14	&	-3.30	&	O{\sc i} 7772 {\AA} (a), P11- P16 (a)	\\ \\
J004927.87+551549.0	&	0.6	&	0.06	&	A0V	&	A2IV	&	13.99	&	13.80	&	-2.77	&	2016-09-19	&	-1.70	&	Mg{\sc ii} 4481 {\AA} (a), Fe{\sc ii} 5169 {\AA} (a), O{\sc i} 7772 {\AA} (a), P11- P13 (a)	\\ \\
J005911.26+555720.1	& 0.8	&	0.11	&	A3III	&	A6IV	&	14.29	&	13.55	&	-2.68	&	2016-09-19	&	-4.84	&	Fe{\sc ii} 5169 {\AA} (a), Fe{\sc ii} 5317 {\AA} (a), O{\sc i} 7772 {\AA} (a), P11- P16 (a)	\\ \\
J011345.53+383819.9	&	0.1	&	0.03	&	A2V	&	A1IV	&	13.47	&	13.10	&	-2.75	&	2015-12-21	&	-0.70	&	Mg{\sc ii} 4481 {\AA} (a), Fe{\sc ii} 5169 {\AA} (a), O{\sc i} 7772 {\AA} (a), P11- P16 (a)	\\ \\
J011452.80+542305.3	&	1.1	&	0.03	&	A0III	&	A3IV	&	12.27	&	11.92	&	-2.80	&	2014-11-04	&	-1.10	&	Mg{\sc ii} 4481 {\AA} (a)  , O{\sc i} 7772 {\AA} (a), P11- P16 (a)	\\ \\
J011650.55+492723.3	&	0.4	&	0.06	&	A0III	&	A2V	&	13.39	&	13.21	&	-2.69	&	2015-12-21	&	-1.50	&	Mg{\sc ii} 4481 {\AA} (a), Fe{\sc ii} 5169 {\AA} (a), O{\sc i} 7772 {\AA} (a), P11- P15 (a)	\\ \\
J012208.07+231347.2	&	0.2	&	0.08	&	A3V	&	A6IV	&	12.11	&	11.72	&	-2.74	&	2012-10-01	&	-0.41	&	Mg{\sc ii} 4481 {\AA} (a), Fe{\sc ii} 5169 {\AA} (e), Fe{\sc ii} 5317 {\AA} (a),\\
&&&&&&&&&&O{\sc i} 7772 {\AA} (a), P11- P17 (a)	\\ \\
J020337.34+284455.1	&	0.2	&	0.11	&	A2III	&	A6IV	&	13.41	&	12.96	&	-2.77	&	2013-10-31	&	-2.80	&	Mg{\sc ii} 4481 {\AA} (a), Fe{\sc ii} 5169 {\AA} (e), O{\sc i} 7772 {\AA} (a), P11- P16 (a)	\\ \\
J021331.68+561900.8	&	1.0	&	0.03	&	A0III	&	A2V	&	14.25	&	13.92	&	-2.65	&	2011-11-13	&	-8.49	&	Fe{\sc ii} 5317 {\AA} (a), O{\sc i} 7772 {\AA} (a), P11- P16 (a)	\\ \\
J021405.28+584915.9	&	1.5	&	0.00	&	A0III	&	A2V	&	14.33	&	13.81	&	-2.74	&	2011-11-13	&	-4.24	&	Mg{\sc ii} 4481 {\AA} (a), O{\sc i} 7772 {\AA} (a), P11- P14 (a)	\\ \\
J030757.57+262544.2	&	0.5	&	-0.03	&	A1V	&	A2V	&	15.00	&	14.88	&	-2.78	&	2011-12-03	&	-1.28	&	P11- P12 (a)	\\
		&		&		&		&	A1V	&		&		&		&	2011-12-11	&	-0.64	&	P11- P15 (a)	\\
		&		&		&		&	A2V	&		&		&		&	2012-01-12	&	--	&	P11- P15 (a)	\\ \\
J031108.31+530719.3	&	1.6	&	0.16	&	A0V	&	A2IV	&	14.26	&	13.23	&	-2.25	&	2011-12-15	&	-3.39	&	Mg{\sc ii} 4481 {\AA} (a), Fe{\sc ii} 5169 {\AA} (e), O{\sc i} 7772 {\AA} (e), P11- P18 (e),\\
&&&&&&&&&&Ca{\sc ii} 8498-8542-8662 {\AA} (e), O{\sc i} 8446 {\AA} (e)	\\ \\
J032758.90+490110.3	&	0.9	&	0.02	&	A1III	&	A3IV	&	11.97	&	11.57	&	-2.72	&	2015-12-21	&	-0.30	&	Mg{\sc ii} 4481 {\AA} (a), Fe{\sc ii} 5169 {\AA} (a), O{\sc i} 7772 {\AA} (a), P11- P17 (a)	\\ \\
J033617.64+585358.3	&	2.7	&	0.03	&	A2V	&	A3IV	&	14.90	&	13.99	&	-2.92	&	2013-09-23	&	-1.63	&	O{\sc i} 7772 {\AA} (a), P11- P16 (a)	\\ \\
J034018.06+580620.4	&	2.0	&	0.03	&	A0III	&	A2V	&	14.61	&	13.82	&	-2.67	&	2013-09-23	&	-6.23	&	Fe{\sc ii} 5169 {\AA} (a), Fe{\sc ii} 5317 {\AA} (e), O{\sc i} 7772 {\AA} (a), P11- P17 (a)	\\ \\
\hline
\end{tabular}
{\footnotesize\begin{flushleft}
(e)-- emission profile; (a)-- absorption profile
\end{flushleft}
}
\end{table}
\end{landscape}

\end{document}